\documentclass[pra,twocolumn,floatfix]{revtex4}
\usepackage{amsfonts,amsmath,amssymb}
\usepackage{graphicx}

\begin{document}

\newlength{\figwidth}
\setlength{\figwidth}{0.45\textwidth}

\preprint{\noindent Prepare for Phys. Rev. A, version 0.1,
\hfill \today} \ \\ \ \\

\title{Correlation dynamics between electrons and ions in the
fragmentation of D$_2$ molecules by short laser pulses}

\author{X. M. Tong,\footnote{Contact: {xmtong@phys.ksu.edu}}
 Z. X. Zhao and C. D. Lin}

\affiliation{J. R. Macdonald Laboratory, Physics Department,
Kansas State University, Manhattan, KS 66506-2601}

\begin{abstract}
We studied the recollision dynamics between the electrons and
D$_2^+$ ions following the tunneling ionization of D$_2$ molecules
in an intense short pulse laser field. The returning electron
collisionally excites the D$_2^+$ ion to excited electronic states
from there D$_2^+$ can dissociate or be further ionized by the
laser field, resulting in D$^+$ + D or D$^+$ + D$^+$,
respectively. We modeled the fragmentation dynamics and calculated
the resulting kinetic energy spectrum of D$^+$ to compare with
recent experiments. Since the recollision time is locked to the
tunneling ionization time which occurs only within fraction of an
optical cycle, the peaks in the D$^+$ kinetic energy spectra
provides a measure of the time when the recollision occurs. This
collision dynamics forms the basis of the molecular clock where
the clock can be read with attosecond precision, as first proposed
by Corkum and coworkers. By analyzing each of the elementary
processes leading to the fragmentation quantitatively, we
identified how the molecular clock is to be read from the measured
kinetic energy spectra of D$^+$ and what laser parameters be used
in order to measure the clock more accurately.
\end{abstract}
 \maketitle

\section{Introduction}
The fragmentation and ionization of D$_2$ by intense optical laser
fields has been an active area of theoretical and experimental
studies during the past decades
\cite{Sandig00,Staudte02,Niikura02,Niikura03,Sakai03}. In most of
these experiments it was assumed that the D$_2$ molecule is
ionized in the early phase of the laser field producing D$_2^+$
ion which is subsequently ionized by the laser. Mechanisms for the
ionization of D$_2^+$ ion include bond softening (SO)
\cite{Posthumus99}, charge resonance enhanced ionization (CREI)
\cite{Codling89,Seideman95,Zuo95a,Constant96,Suzor95,Codling93,Bandrauk99b},
in addition to  direct ionization by the laser field. The
dissociation and ionization of D$_2^+$ in the laser field result
in D$^+$ + D or D$^+$ + D$^+$, with characteristic kinetic
energies reflecting the internuclear separation of the breakup of
D$_2^+$ at the time when it is excited or ionized. Thus bond
softening and CREI, which produce distinct peaks in the D$^+$ ion
kinetic energy spectra have been observed experimentally and
predicted theoretically. These peaks can be understood without
reference to the ionization of D$_2$ itself initially, i.e., the
ionization of D$_2$ and D$_2^+$ can be treated as two independent
events. However, recent experiments
\cite{Staudte02,Niikura02,Niikura03,Sakai03} pointed out a new
group of peaks in the D$^+$ ion spectra at the higher energy
(about 5 eV to 10 eV per ion) which has now been attributed to the
rescattering process \cite{Fittinghoff92,Kondo93a,Corkum93,Walker94a,Brabec96,%
Sheey98,Becker00,Kopold00a,Yudin01,Yudin01a,Fu01}. In the
rescattering process, the electron which is released by tunneling
ionization is driven back by the laser field to collide with the
residual D$_2^+$ ion to ionize it or to excite it. If the D$_2^+$
ion is excited, it can dissociate directly or be further ionized
by the laser. In both cases, the D$^+$ ion will have kinetic
energy (the reflection principle) characteristic of the
internuclear separation where the ionization occurs. This paper is
to examine all the elementary processes that lead to the emission
of D$^+$ ions by the rescattering process following the initial
tunneling ionization of the D$_2$ molecule. The calculation will
be performed for D$_2$ molecules only, but clearly the same model
can be applied to H$_2$ with minor modifications.

In Sec.~\ref{sec:theory} we first discuss all the elementary
processes that lead to the dissociation or ionization of D$_2$
molecules in a laser field. Starting with the tunneling ionization
of D$_2$, we address the following issues: (1) Calculation of the
ionization rates of D$_2$ from its equilibrium distance using the
molecular tunneling ionization (MO-ADK) theory \cite{Tong02b}; (2)
The classical trajectory of the ionized electron in the laser
field and the Coulomb field of the D$_2^+$ ion, with initial
longitudinal and transverse velocity distributions following the
description of the ADK theory \cite{Yudin01a}; (3) The free
propagation and spreading of the nuclear wave packet after the
tunneling ionization of D$_2$ from its equilibrium distance; (4)
Semi-empirical formulae for electron impact excitation cross
sections of D$_2^+$ from the $\sigma_g$ ground state to the first
few excited electronic states, in particular, the first $\sigma_u$
and $\pi_u$ states. These cross sections have to be evaluated at
all values of internuclear separations and for different alignment
angles of D$_2^+$ with respect to the laser polarization
direction; (5) Evaluation of tunneling ionization rates of D$_2^+$
from the excited $\sigma_u$ and $\pi_u$ states at each
internuclear separation; (6) Follow the time evolution of
dissociation and ionization dynamics to extract the kinetic energy
spectra of the fragmentation products. While rates or cross
sections for each of these elementary processes have been
formulated, in the full calculation we only consider D$_2$
initially aligned perpendicular to the direction of the laser
polarization. In Sec.~\ref{sec:results}, the resulting kinetic
energy spectra of D$^+$ are compared to the experiment of Niikura
{\it et~al.} \cite{Niikura03} where D$^+$ ions were detected
without knowing whether the other fragmentation product is a D or
a D$^+$. In contradiction to the conclusion of this work where the
main peak in the D$^+$ kinetic energy spectrum was attributed to
the dissociation of D$_2^+$ following excitation to the $\sigma_u$
electronic state, we conclude from our calculation that the main
peak is due to the  further ionization  of the excited D$_2^+$ by
the laser. This has the consequence that the molecular clock we
read is at a different time from the one read in Niikura {\it
et~al.} \cite{Niikura03}. We further analyzed the contributions of
the total kinetic energy spectra of D$^+$ resulting from the
different excited electronic states, from dissociation or
ionization, and from rescattering at the first return, the third
return, or higher returns. The simulated D$^+$ kinetic energy
spectra from ionization also were compared to the recent
experiment of Alnaser {\it et~al.} \cite{Alnaser03}. From our
analysis, we conclude that the fragmentation of D$_2$ can be used
as a molecular clock based on the rescattering dynamics. The clock
can be read more accurately if the laser pulse is chosen at the
lower intensity and with a shorter duration. With such a clock,
the time duration can be read with an accuracy of fraction of a
femtosecond without the attosecond laser pulses. Equivalently this
means  that the distance between the two nuclei can be read with
an accuracy of fractions of an Angstrom. This can be achieved
experimentally by comparing kinetic energy spectra for experiments
carried out at different mean laser wavelengths, or by comparing
the kinetic energy spectra of D$^+$ and H$^+$ from the
fragmentation of D$_2$ and H$_2$, respectively, in the same laser
pulse. We finish this paper with a summary and conclusion in
Sec.~\ref{sec:conclusion}.

\section{The Theoretical Model}\label{sec:theory}

\subsection{The elementary processes}
\begin{figure}[tb]
 \includegraphics[width=1.15\figwidth,height=3.5in]{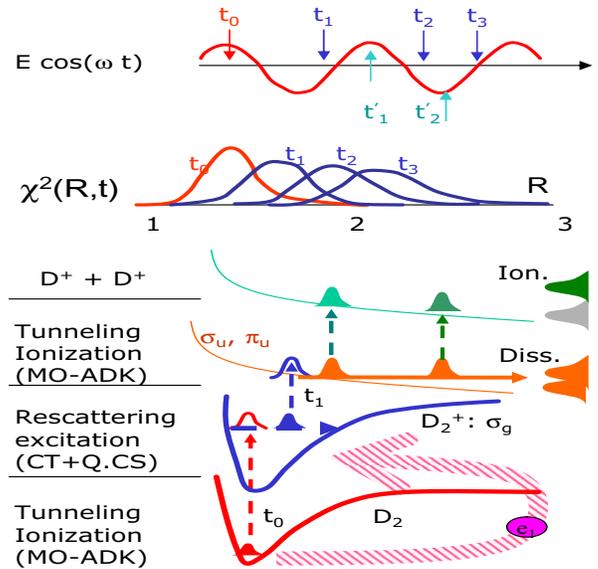}
 \caption{\label{fig:sch}
 Schematic of the major physical processes leading to the
 formation of D$^+$ ion by the dissociation
 or ionization of D$_2^+$. The first row depicts the
 oscillating electric field of the laser. The second row shows
 the spreading of the vibrational wave packet after the initial
 tunneling ionization. At t$_0$ the D$_2$ molecule is ionized.
 The tunneled electron returns to rescatter with the D$_2^+$
 ion at t$_1$ where it excites D$_2^+$ to the excited
 $\sigma_u$ or $\pi_u$ electronic states. The excited D$_2^+$
 can dissociate along these repulsive potential curves, or
 further ionized by the laser to produce D$^+$ ions by Coulomb
 explosion, to produce characteristic kinetic energy
 spectra of the fragments. For more detailed description, see Text.}
\end{figure}

The schematic of the physical processes leading to the
fragmentation of D$_2^+$ ion  following the ionization of D$_2$
molecule in an intense laser pulse is depicted in
Fig.~\ref{fig:sch}.
The D$_2$ molecule is first ionized at t$_0$ near the peak of the
laser pulse, releasing an electron into the oscillating laser
field. At t$_1$, the electron is driven back to the molecular ion,
to excite the other electron in the ion to one of the higher
excited electronic states, or to ionize it. We will be dealing
with peak laser intensities such that the returning electron does
not have enough energy to ionize D$_2^+$.  In the meanwhile, the
nuclear wave packet propagates from its mean internuclear distance
R$_0$=1.4 a.u. at t$_0$ to R$_1$=1.6 a.u at t$_1$.   Thus the
electron impact excitation probabilities of the molecular ion by
the returning electron have to be calculated for D$_2^+$ with a
vibrational distribution $\chi^2(R, t_1)$. This distribution is
indicated in the second row of Fig.~\ref{fig:sch}. Once the
D$_2^+$ is in the excited state, represented by the curve labelled
$\sigma_u$ and $\pi_u$  in Fig.~\ref{fig:sch}, the D$_2^+$ can
dissociate directly to D$^+$ + D, or it can be further ionized by
the laser at t$'_1$ when the electric field of the laser returns
to its peak value. If the D$_2^+$ is ionized at t$'_1$, then it
will fragment by Coulomb explosion to produce D$^+$ + D$^+$ ions.
The total released kinetic energy for such a ``two-step'' process
can be calculated.

The rescattering does not have to occur only at the first return
time t$_1$. Due to the attractive field from the molecular ion,
the released electron can return to collide with the molecular ion
at later times, i.e., after more than one optical cycle, following
the initial ionization. For example, the return can occur at t$_2$
and  t$_3$, in the second optical cycle, or at t$_4$ and t$_5$, in
the third optical cycle, and so on. At these later times
excitation and ionization   occur at larger internuclear
separations, thus the kinetic energies of the fragmented D$^+$ ion
are smaller. In general the returning probabilities become small
after three optical cycles. Following the general convention we
call t$_2$ the second return and t$_3$ the third return, etc...

An important feature of the elementary processes described above
is that the rescattering times t$_i$ and the subsequent tunneling
ionization time t$'_i$ are relatively well locked to the clock
t$_0$ of the initial ionization of D$_2$. Since tunneling
ionization occurs only near the peak of the laser field, t$_0$
spans only a fraction of an optical cycle. Similarly, t$_i$ and
t$'_i$ are also restricted to within sub-fs accuracy. These
precise clocks in turn define precise internuclear separations.
For laser pulses with mean wavelength at 800 nm, the mean
internuclear distances R$_i$ for t$_i$ (i=1,3,5,7) are shown in
Table~\ref{tab:return} for H$_2$ and D$_2$. Note that at t$_7$,
the center of the vibrational wave packet for H$_2^+$ has already
bounced back from the outer turning point, but not so for D$_2^+$.
A classical estimate shows that it takes 8.5 fs to reach its outer
turning point. During these later times, the wave packet spreads
significantly.
\begin{table}[tb]
\caption{\label{tab:return} Relation between the returning time
and the average nuclear separation for H$_2^+$ and D$_2^+$.}
\begin{ruledtabular}
\begin{tabular}{||c|c|c|c||}
    return   & time (fs) & \multicolumn{2}{c||}
    {$<R>$ (a.u.)} \\ \hline
              &        & H$_2^+$ & D$_2^+$ \\ \hline
    t$_1$     & 1.9    & 1.8 & 1.6 \\
    t$_3$     & 4.3    & 2.5 & 2.1 \\
    t$_5$     & 7.0    & 3.0 & 2.6 \\
    t$_7$     & 9.6    & 3.2 & 3.0 \\
\end{tabular}
\end{ruledtabular}
\end{table}
The clock or the mean internuclear separation can be probed
directly by the characteristic kinetic energy peaks of the
fragmented D$^+$ ions. Changing the wavelength of the laser
clearly will change the clocks and the mean internuclear
separations. Replacing D$_2$ by H$_2$ will not change the clock
but will change the mean internuclear distances.

To read the clock from the measured kinetic energy of the
fragmented ion, however, there are a number of factors that make
the clock ``fuzzy''. First, the initial tunneling ionization
occurs over an interval of about 0.3 fs near the peak of the laser
field. The initial vibrational wave packet, taken to be the ground
vibrational wavefunction of D$_2$, according to the Frank-Condon
principle, has a width of 0.2 nm. This vibrational wave packet
will broaden as it expands to larger internuclear separation. The
electron impact excitation probabilities and the MO-ADK rates also
depend on  internuclear separations. These factors would reduce
the precision of the clock such that distinct peaks in the kinetic
energy distribution of the fragmented ions are not as clearly
separated. We model the rescattering process to check how
accurately the molecular clock can be read from the kinetic energy
spectra of the fragmented ions for different laser parameters.

We now describe the models used for calculating the rates and
probabilities for each elementary process.

\subsection{Tunneling ionization rates for molecules}

We first discuss how the ionization rates of D$_2$ and D$_2^+$ in
the laser fields are calculated. From the rescattering model
above, we need the ionization rates for D$_2$ from the ground
state, and for D$_2^+$ from the excited $\sigma_u$ and $\pi_u$
states over the whole range of R.   The rates are needed for
different alignment of the molecules as well.

We calculated the tunneling ionization rates using the recently
developed MO-ADK model \cite{Tong02b}. It was obtained by
extending the widely used ADK \cite{Perelomov66,Ammosov86} theory
for atoms in a laser field to molecules. In the MO-ADK theory the
ionization rates are given in semi-analytical expressions. For a
diatomic molecule in a parallel static electric field, the
ionization rate for a valence electron is given by
\begin{eqnarray}\label{eq:staticR}
W_m(F) & = &   \frac{B^2(m)}{2^{|m|} |m|!}
\frac{1}{\kappa^{2Z_c/\kappa-1}} \nonumber \\ & & \times\
\left(\frac{2\kappa^3}{F}\right)^{2Z_c/\kappa - |m|-1}
 e^{-2\kappa^3/3F},
\end{eqnarray} with
\begin{eqnarray}\label{eq:Bm}
B(m) = \sum_l  C_{lm}(-1)^m \sqrt{\frac{(2 l +1) (l+|m|)!}{2
(l-|m|)!}}.
\end{eqnarray}
Atomic units are used unless otherwise indicated. In
Eq.~(\ref{eq:staticR}), $\kappa$ is related to the ionization
energy I$_p$ by $\kappa = \sqrt{2 I_p}$, $l$ is the orbital
angular momentum of the valence electron, $m$ is its projection
along the internuclear axis, Z$_c$ is the effective charge seen by
the valence electron in the asymptotic region and  $F$ is the
field strength. In Eq.~(\ref{eq:Bm}), the parameters C$_{lm}$ are
determined from the valence electron wave function of the molecule
in the asymptotic region. The laser peak power will be given in
units of I$_0 = 10^{14}$ W/cm$^2$ and the mean wavelength is 800
nm. If the molecule is aligned at an angle $\theta$ with respect
to the laser polarization direction, the ionization rate is given
by
\begin{eqnarray}\label{eq:aR}
W_m(F,\theta) & = &  \sum_{m'} W_{m'}(F),
\end{eqnarray}
where $W_{m'}$ is given in Eq.~(\ref{eq:staticR}) except that
\begin{eqnarray}\label{eq:aC}
B(m')  & = & \sum_l  C_{lm} D^l_{m',m}(0,\theta,0)\nonumber \\ & &
\times (-1)^{m'} \sqrt{\frac{(2 l +1) (l+|m'|)!}{2 (l-|m'|)!}},
\end{eqnarray}
where the $D$-function expresses the rotation of the electronic
wave function from the direction of the molecular axis to the
laser polarization direction.  In the MO-ADK model,
Eq.~(\ref{eq:staticR}) reduces to the traditional ADK model for
atoms if $l$ is taken to be the orbital angular momentum quantum
number of the valence electron. For diatomic molecules, the
summation over $l$ is a consequence of expanding the two-center
electronic wave function  in terms of single-center atomic
orbitals. The coefficients C$_{lm}$ are functions of R and depend
on the electronic states of the molecule.

For D$_2$ at the equilibrium internuclear separation, the
parameters C$_{lm}$ have been calculated by Tong {\it et~al.}
\cite{Tong02b}.  Within the range of its ground vibrational
wavefunction, it was found  that the MO-ADK rates  depend  weakly
on R. It was also found that the  major component of B(m=0) in
Eq.~(\ref{eq:Bm}) is $l$=0, for D$_2$, thus the MO-ADK rates for
D$_2$ depend weakly on the alignment of the molecule. The accuracy
of the MO-ADK rates for D$_2$ at the equilibrium distance had been
checked previously and found  to be in good agreement with the
result from {\it ab initio }  calculations \cite{Saenz00a}.

\begin{figure}[tb]
 \includegraphics[width=\figwidth]{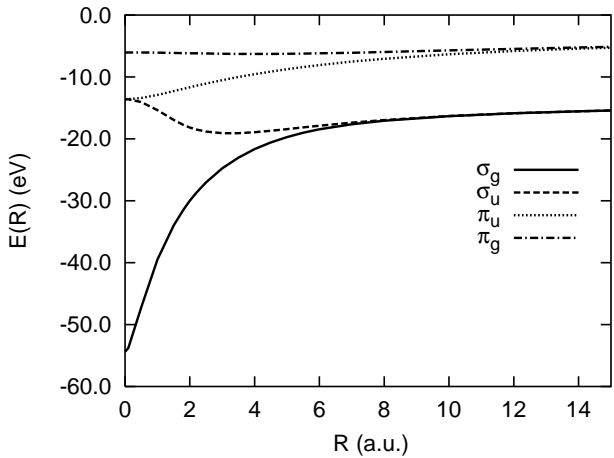}
 \caption{\label{fig:pot}
 Binding energies of D$_2^+$ as a function of internuclear separation.}
\end{figure}

We next consider the ionization of D$_2^+$ in a laser field. Since
the ionization rate depends sensitively on the ionization
potential, in Fig.~\ref{fig:pot}
we show the electronic {\em binding energies}\  E$_i$(R) at each R
of the first four electronic states of D$_2^+$. The negative of
the electronic binding energy is the ionization potential. The
total potential energy  of each electronic state is U$_i$(R)=
E$_i$(R)+ 1/R. For peak laser intensity in the range of 0.5-5
I$_0$, estimate based on the simple ADK theory or the more
complete MO-ADK theory shows that D$_2^+$ in the $\sigma_g$ state
will not be ionized by the laser except for R greater than about 5
a.u., while for $\pi_u$ and $\pi_g$ states the D$_2^+$ will be
readily ionized because of the much smaller ionization potentials.
Thus we need to calculate only the MO-ADK rates of D$_2^+$ in the
$\sigma_u$ state as a function of the internuclear distance R.

\begin{figure}[b]
 \includegraphics[width=\figwidth]{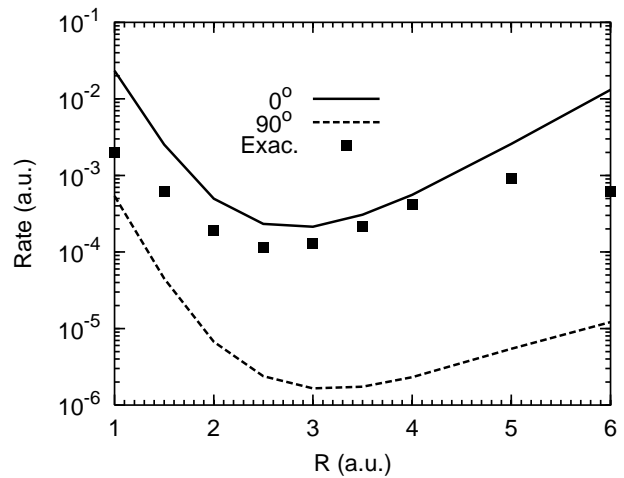}
 \caption{\label{fig:adk}
 Static MO-ADK ionization rates  for D$_2^+$  molecules
 aligned parallel (solid curve) and perpendicular (dashed curve)
 to the electric field direction. The filled squares represent
 exact static ionization rates calculated using the complex
 rotation method \cite{Chu01a} for parallel aligned molecules.
 The static field strength is 0.06 a.u..}
\end{figure}

In Fig.~\ref{fig:adk}
we show the calculated MO-ADK tunneling ionization rates at F=0.06
a.u. for the $\sigma_u$ electronic state of D$_2^+$, for R in the
range of 1-6 a.u. and for alignment angle $\theta$=0$^\circ$ and
90$^\circ$. At $\theta$=0$^\circ$, the MO-ADK rates have been
checked against the ``exact'' static tunneling ionization rates
calculated using the complex rotation method in the two-center
system \cite{Chu01a}. The MO-ADK rates tend to be somewhat higher,
especially at small and large R region. For R greater than 6.0 the
$\sigma_u$ ionization energy is  already very close to the
ionization energy of atomic H, thus  the ADK ionization rates of
H(1s) are used  for $R > 6.0$.  In the actual calculation, the
coefficients C$_{lm}$ are obtained for each R such that the MO-ADK
rates can be readily calculated for any field strength, and any
alignment angle of the molecule using Eq.~(\ref{eq:aR}).

\subsection{The rescattering model}
Following the initial ionization of D$_2$, a correlated electron
wave packet and a vibrational wave packet are created at t$_0$.
The initial vibrational wave packet is taken to be the ground
vibrational wavefunction of D$_2$, assuming that the ionization
process is fast and the Frank-Condon principle is valid. Due to
the heavy mass of the nuclei, the vibrational motion is not
modified by the subsequent laser field. The time evolution of the
vibrational wave packet is thus described by
\begin{eqnarray}
\chi(R,t) & = & \sum_v C_v \chi_v(R) e^{-i \varepsilon_v t}, \\
C_v & = & \int \chi_g(R) \chi_v(R) dR.
\end{eqnarray}
Here \{$\chi_v(R)$\} and \{$\varepsilon_v$\} are  the vibrational
wavefunction and vibrational energy of D$_2^+$ in the $\sigma_g$
ground electronic state, respectively, and $\chi_g(R)$ is the the
ground vibrational wave function of D$_2$.

The rescattering model for describing the motion of the ionized
electron in the subsequent laser field is modeled similar to the
method used by Yudin and Ivanov \cite{Yudin01,Yudin01a} for He.
The ionized electron is treated classically, under the combined
force from the laser field and the residual Coulomb interaction
from the D$_2^+$ ion. For simplicity, the latter is approximated
by an effective charge Z$_c$=+1 at the midpoint of the
internuclear axis. To calculate the trajectory of the ionized
electron, we solve the equation of motion (Newton's second law),
with the initial condition that the ionized electron is at
(x,y,z)=(0,0,z$_0$) where z$_0$ is the tunneling position from the
combined potential of the Coulomb field and the static electric
field. The initial velocity $\mathbf{v}$ is assumed to have a
distribution from the ADK model,
 \begin{eqnarray}\label{eq:v}
g(\mathbf{v}) &\propto & e^{-\mathbf{v}^2 \kappa / F}.
\end{eqnarray}
In this model, the tunneled electron is ejected isotropically with
a Gaussian distribution in velocity, i.e., we consider the ejected
electron have initial velocity in both the transverse and the
longitudinal directions. For each initial time t$_0$ or phase
$\phi_0$ that the ionized electron was born, the classical
equation of motion was solved to obtain the trajectory. The
distance of the electron from the center of D$_2^+$ ion is
monitored for over seven optical cycles for longer pulses or till
the end of the laser pulse if the pulse is shorter. The distance
of closest approach of the electron from the ion and the time when
this occurs for each trajectory are recorded. From these data, the
impact parameter b and the collision energy T of the corresponding
electron-ion impact (no laser field) excitation or ionization are
obtained.

\begin{figure}[tb]
 \includegraphics[width=\figwidth]{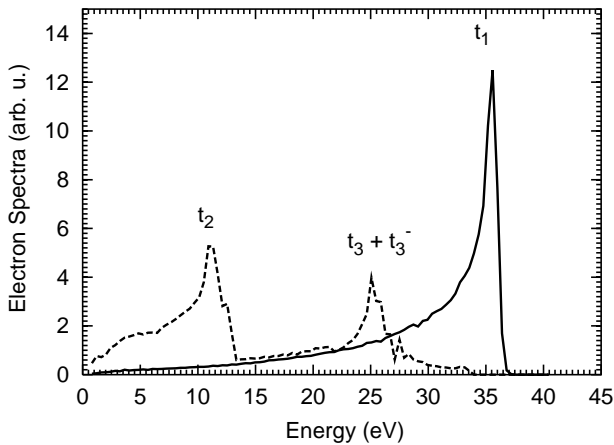}
 \caption{\label{fig:eng} Returning electron energy spectra for D$_2$
 in a pulse laser with peak intensity of 1.5 I$_0$ (I$_0$=10$^{14}$ W/cm$^2$)
 and pulse length of 40 fs obtained from the simulation.
 }
\end{figure}

Figure~\ref{fig:eng} shows the probability distribution of finding
the returning electron with kinetic energy T measured in the
asymptotic region for a laser with peak intensity at 1.5 I$_0$. If
the residual Coulomb interaction from the D$_2^+$ ion is
neglected, the expected maximum returning energy will be 3.17
U$_P$ = 29 eV, where U$_P$ is the Pondermotive energy. The
inclusion of Coulomb interaction increases this peak energy to
about 35 eV.

In Fig.~\ref{fig:eng} we show three groups of returning electrons.
In the first group, the electron was born at t$_0$ when the laser
field has a positive phase $\phi_0$ (i.e., beyond the peak field).
It was driven outward and then back by the oscillating laser field
to recollide with the D$_2^+$ ion within one optical cycle. This
group is denoted by t$_1$ where the returning  electron has peak
current near 35 eV. The second group  labelled t$_2$ denotes an
electron which does not collide with the ion at the first return,
but at the second return about  half a cycle later after the
electron reverse its direction again.  The kinetic energy for this
group of electrons is smaller. The  third group was denoted by
t$_3$+t$_3^-$. For the t$_3$, the  recollision occurs at the third
return. For the t$_3^-$ group, the electrons were born at a
negative phase $\phi_0$ \cite{Fu01}, i.e., before the laser
reaches the peak field. These negative phase electrons do not
recollide with the ion in the first optical cycle when the field
change direction since they were accelerated by an increasing
field right after  birth. Due to the Coulomb focusing by the ion
they collide with the ion at  the third return. Without the
Coulomb focusing the negative  phase birth would not contribute to
the rescattering process. In calculating the returning electron
energy distribution shown in Fig.~\ref{fig:eng} proper weights
from the MO-ADK rates and the initial velocity distribution of the
tunneling electron have been accounted for. In Fig.~\ref{fig:eng}
we did not show the electron energy distributions from collisions
occurred at returns after two optical cycles. The general trend is
that at higher returns,   the kinetic energy of the electron is
smaller and the probability of rescattering is also smaller. In
our calculations we have accounted for rescattering up to seven
optical cycles for the long laser pulses.

\subsection{Electron impact excitation and ionization probabilities}
For each impact parameter b and kinetic energy T of the returning
electron, we need to calculate the electron impact excitation and
ionization cross sections of D$_2^+$ at each internuclear
separation R. Different from the He$^+$ case, there are few
experimental or theoretical data available for D$_2^+$. Thus we
have to generate the cross sections needed semi-empirically. For
each total cross section $\sigma(T)$ at kinetic energy T, we
assume that the probability for excitation or ionization at impact
parameter b is given by
 \begin{eqnarray}\label{eq:improb}
    P_{m} (b,T) & = & \sigma(T) \frac{e^{-b^2/a_o^2}}{\pi a_o^2}, \\
    a_o & = & \sqrt{2 /\Delta E},
\end{eqnarray}
where $T=v^2/2$ and $\Delta E$ is the excitation or ionization
energy. Here, the b-dependence is taken to be the Gaussian form.
For the rescattering in He, Yudin and Ivanov \cite{Yudin01} have
checked different forms of b-dependence and concluded that the
results are rather insensitive to the precise functional form
used.

For electron impact ionization cross section, we employ the
empirical formula
 \begin{eqnarray}
\sigma_i(T,\Delta E) & = & \frac{\pi}{\Delta E^2} e^{1.5*(\Delta
 E-0.5)/T}f(T/\Delta E) \label{eq:si} \\
f(x) & = & (A \ln{x} + B(1-\frac{1}{x})-C
\frac{\ln{x}}{x})\frac{1}{x} \label{eq:fx}.
\end{eqnarray}
where $\Delta E$ is the ionization energy. By fitting this formula
to the accurate theoretical H(1s) ionization cross section
\cite{Bray} we obtained A = 0.7213, B=-0,302, C=0.225. The fitted
formula, when applied to He$^+$, gives ionization cross sections
in good agreement with the theoretical results of Bray \cite{Bray}
for He$^+$ as well. For D$_2^+$ at the equilibrium distance this
formula also reproduces the recommended ionization cross section
from NIST \cite{Kim00}. In this semi-empirical model, the
molecular ion is treated as a point particle, thus the ionization
cross section is independent of the alignment of the D$_2^+$ ion.

\begin{figure}[tb]
 \includegraphics[width=\figwidth]{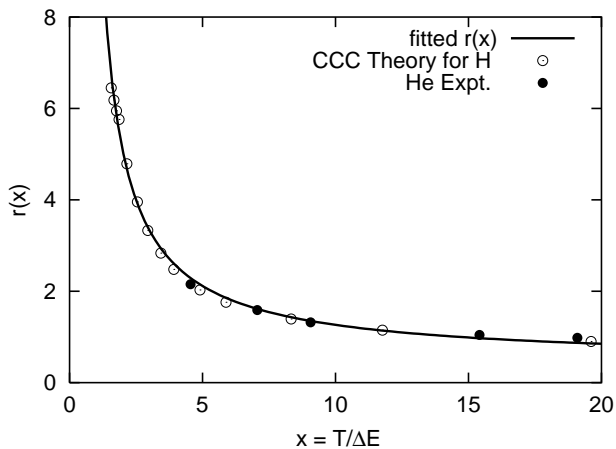}
 \caption{\label{fig:ratio} Ratio of the electron impact excitation cross
 section to 2p$_0$ with respect to 2p$_1$ vs scaled excitation energy.
 Solid line is the fitted result from Eq.~(\ref{eq:rx}), open circles are from
 the  calculation of Bray for H \cite{Bray03}, and filled circles are from the He
 experimental measurement \cite{Merabet99}.}
\end{figure}

For the excitation process, it is clear from Fig.~\ref{fig:pot}
that $\sigma_u$ and $\pi_u$ states will be the dominant channels
populated via electron impact excitation from the ground
$\sigma_g$ state since they have the lowest excitation energies.
There are no theoretical or experimental data available for such
cross sections as functions of internuclear separations. Thus we
will employ semi-empirical fitting procedure as well. We assume
that the excitation cross section again can be fitted in the form
of Eqs (\ref{eq:si}) and (\ref{eq:fx}) as in ionization, except
that $\Delta E$ now is the excitation energy and the number 0.5 in
Eq (\ref{eq:si}) should be replaced by the excitation energy of
the corresponding state in atomic hydrogen. From the tabulated
H(1s)$\to$ H(2p) excitation cross section in Bray \cite{Bray}, we
obtained A = 0.7638, B=-1,1759, C=-0.6706. The formula was further
tested by comparing the predicted excitation cross section with
the calculated one for e$^-$+ He$^{+}$(1s) $\rightarrow$ e$^-$+
He$^{+}$(2p). From the total 1s $\to$ 2p excitation cross section,
we can further distinguish excitation cross section to 2p$_0$ or
2p$_1$, with the direction of the incident electron beam as the
quantization axis. The relative 2p$_0$ and 2p$_1$ cross sections
can be calculated theoretically or experimentally from
polarization or correlation measurements. (Note: 2p$_{-1}$ cross
section is identical to 2p$_1$ cross section by symmetry.) In
Fig.~\ref{fig:ratio} we show the relative cross sections of
2p$_0$ to 2p$_1$ from the calculation of Bray \cite{Bray03} for H,
plotted against scaled energy (with respect to the excitation
energy). On the same graph we display the same ratio for the
excitation of He from 1s$^2$ to 1s2p$^1P^o$ from the experiment of
Merabet {\it et~al.} \cite{Merabet99}. It appears that the both H
and He data fall on the same curve when the collision energy is
scaled with respect to the excitation energy. We fit the 2p$_0$ to
2p$_1$ cross section ratio by
\begin{eqnarray}\label{eq:rx}
r(x) = \frac{\sigma_0}{\sigma_1} = \frac{8.2\sqrt{1+1.1/x^2}}{x} +
0.44 \ .
\end{eqnarray}
where $x$=T/$\Delta E$ is the scaled kinetic energy. Since the
ratio for He does not differ much from the calculated ratio for H,
this comparison convinces us to use the $r(x)$ in
Eq.~(\ref{eq:rx}) to describe the ratio for D$_2^+$  as well. The
$r(x)$ indicates that m=0 is the dominant magnetic component in
the present interested energy regime.

 To relate the 2p$_0$ or 2p$_1$ partial cross sections to the
excitation cross sections of $\sigma_u$ and $\pi_u$ electronic
states of D$_2^+$, we need to know the alignment angle of the
molecule. If the molecule is aligned along the laser field
polarization direction (which is also the direction of the
electron beam), the 2p$_0$ cross section is the excitation to
$\sigma_u$ state and the 2p$_1$ (2p$_{-1}$) cross section is for
the excitation to the $\pi_u$ state. If the molecule is aligned
perpendicular to the laser polarization direction, then the role
is reversed, i.e., 2p$_1$ (or 2p$_{-1}$) corresponds to the cross
section of the $\sigma_u$ excitation, and 2p$_0$ cross section to
the $\pi_u$ excitation. For any arbitrary alignment angle $\theta$
of D$_2^+$, we assume the total excitation cross sections to
$\sigma_u$ and $\pi_u$ are given by
\begin{eqnarray}
\sigma(\sigma_u) & = & \sigma_T (r_0 \cos^2\theta + r_1
\sin^2\theta) \\
\sigma(\pi_u)   & = & \sigma_T (r_0 \sin^2\theta + r_1 \cos^2\theta).\\
\sigma_T & = & \sigma_0 + 2 \sigma_1 \\
r_0 & = & \frac{\sigma_0}{\sigma_T} = \frac{r(x)}{r(x)+2} \\
r_1 & = & \frac{2 \sigma_1}{\sigma_T} = \frac{2}{r(x)+2}.
\end{eqnarray}
The semi-emipirically fitted electron impact ionization or
excitation cross section formulae discussed so far are for a free
electron colliding with  an atomic or molecular ion. For the
rescattering process, the  two electrons in D$_2$ initially are in
the singlet state (S=0).  Thus in principle, one should just use
singlet excitation or ionization  cross sections, instead of the
spin-averaged cross sections. We  obtain the singlet cross
sections from the total cross section following the empirical
formula derived in Yudin and Ivanov \cite{Yudin01a} [their Eqns.
(8) and (9)].

\begin{figure}[tb]
 \includegraphics[width=\figwidth]{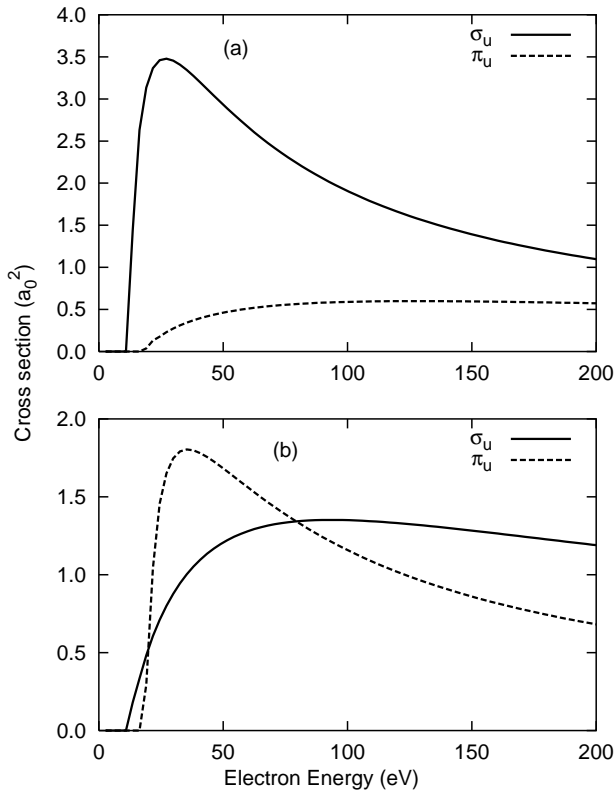}
 \caption{\label{fig:cross} Electron impact excitation cross
 sections to $\sigma_u$ and $\pi_u$ states of D$_2^+$ at the equilibrium distance.
 (a) The electron beam is
 parallel to the molecular axis; (b) the electron beam is
 perpendicular to the molecular axis.}
\end{figure}

 These empirical formulae allow us to calculate  electron
impact excitation cross sections from  $\sigma_g$ to $\sigma_u$
and to $\pi_u$ states at each internuclear separation and at each
alignment of the D$_2^+$ ion. In Fig.~\ref{fig:cross} we compare
the electron
impact excitation cross sections at the equilibrium distance to
$\sigma_u$ and $\pi_u$ states, for D$_2^+$ ions lying parallel and
perpendicular to the incident electron direction which is also the
direction of the laser polarization, respectively. When  D$_2^+$
is aligned parallel to the laser polarization, impact excitation
to $\sigma_u$ is the dominant channel. The $\pi_u$ cross sections
are smaller due to two factors: (1) the $\pi_u$ state has higher
excitation energy, see Fig.~\ref{fig:pot}; (2) the 2p$_0$ state
has larger cross sections than  2p$_1$ for the electron energies
considered, see Fig.~\ref{fig:ratio}. The situation is different
when D$_2^+$ ion is aligned perpendicular to the laser
polarization direction. Fig.~\ref{fig:cross}(b) indicates that
excitation to the $\pi_u$ state is actually larger than that to
the $\sigma_u$ state, at least in the 20-80 eV energy region. Note
that in the experiments of Niikura {\it et~al.}
\cite{Niikura02,Niikura03} the H$_2^+$ or D$_2^+$ were chosen to
be perpendicular to the laser polarization direction. They assumed
that electron impact excitation populates only the $\sigma_u$
state, in disagreement with our analysis.

The semi-empirical formulae presented above allow us to calculate
electron impact excitation cross sections to $\sigma_u$ and
$\pi_u$ states averaged over the initially randomly distributed
D$_2^+$ ions. We obtained the ratio of the cross section of
$\sigma_u$ with respect to $\pi_u$, and compared the result with
the ratio obtained by Peek \cite{Peek64} where the impact
excitation cross sections for different internuclear separations
were calculated using the Born approximation. The agreement is
quite good, with the average cross section for $\sigma_u$ about a
factor of two larger than for $\pi_u$. The absolute cross sections
from Peek are larger since Born approximation was used.

We also consider the small contribution from excitation to the
2s$\sigma_g$ electronic state of D$_2^+$. The empirical formula is
chosen to be
\begin{eqnarray} \sigma_e(T,\Delta E) &
= & \frac{1}{\Delta E^2}
f(T/\Delta E) \\
f(x) & = & \frac{A}{1+B/x} \frac{1}{x}
\end{eqnarray}
where the parameters  A = 0.17, B=1.53 are obtained by fitting the
formula to the 1s $\to$ 2s excitation cross sections of H. This
cross section is assumed to be independent of the alignment of the
molecular ion.

\subsection{Impact excitation probability} 

\begin{figure}[b]
 \includegraphics[width=\figwidth]{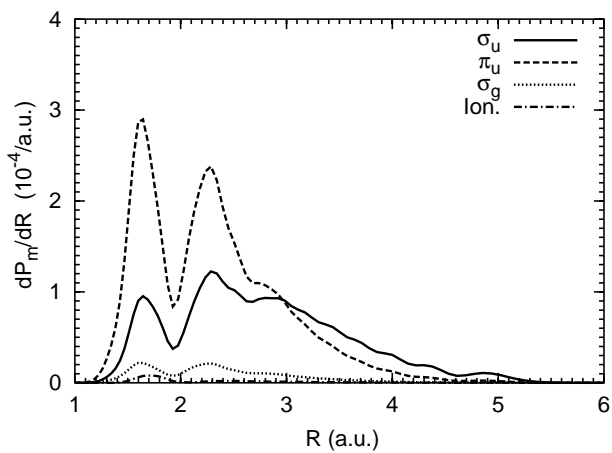}
 \caption{\label{fig:impro} Electron impact excitation
 and ionization probabilities of D$_2^+$ by the rescattering
 electron following tunneling ionizaiton of D$_2$ by a short pulse laser
 with peak intensity of
1.5 I$_0$ (I$_0$=10$^{14}$ W/cm$^2$) and pulse length of 40 fs.}
\end{figure}

With all the elementary cross sections available, we can now
calculate the probability distribution of exciting D$_2^+$ at a
given internuclear separation R from the ground $\sigma_g$ state
to a specific excited electronic state or to ionization states by
the returning electron where the returning electron originates
from the ionization of D$_2$ molecule by the laser over a half
optical cycle. The probability distribution is given by
{\small
\begin{eqnarray}\label{eq:pim}
\frac{d P_{m}}{d R} =  \frac{ \int\int P_m(b,T) \chi^2(R,t_r)
g(\mathbf{v}) W(F\cos\phi) d\mathbf{v} d\phi}{\int\int
g(\mathbf{v}) W(F\cos\phi) d\mathbf{v} d\phi},
\end{eqnarray} }
The subscript $m$ stands for the excited states ($\sigma_u,\pi_u,
\sigma_g$) or ionization. $P_m(b,T)$ is the impact excitation or
ionization probability from Eq.~(\ref{eq:improb}). In this
expression, W is the MO-ADK rate for ionizing D$_2$ at the static
field F$\cos\phi$, where F is the peak field strength of the
laser. For each $\phi$, the tunneled electron leaves the molecule
with an initial velocity $\mathbf{v}$, with a distribution
governed by Eq.~(\ref{eq:v}), i.e., effects due to both the
longitudinal and transverse velocity distributions are included.
For each initial velocity and initial position of the tunneled
electron, the return time t$_r$ at the distance of closest
approach, the corresponding laser-free impact parameter b and
kinetic energy T are calculated, and the excitation probability is
also calculated. At each return time t$_r$, the distribution of
the vibrational wave packet, $\chi^2$(R,t$_r$), is used to
calculate the probability of finding D$_2^+$ at internuclear
separation R. In this expression the MO-ADK rates and the impact
excitation probabilities to $\sigma_u$ and $\pi_u$ states depend
on the alignment of molecules. The other quantities are isotropic.
For D$_2$ initially aligned perpendicular to the direction of the
linear polarization of the laser, the impact excitation
probabilities at different R's over half an optical cycle are
shown in Fig.~\ref{fig:impro}, where the peak laser intensity is
1.5 I$_0$. Note that
 excitation probability to $\pi_u$ is the largest, but
to $\sigma_u$  is also significant. On the other hand, excitation
to  2s$\sigma_g$ excited state  and direct ionization by the
rescattering electron are not important.

\begin{figure}[tb]
 \includegraphics[width=\figwidth]{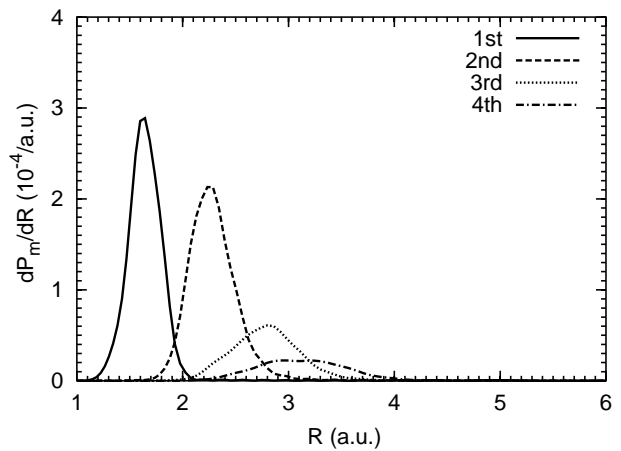}
 \caption{\label{fig:impro2} Electron impact
 excitation probabilities by the rescattering electron to the
 $\pi_u$ state in the first four optical cycles after D$_2$ molecules are
 ionized by a short pulse laser of peak intensity of
  1.5 I$_0$ (I$_0$=10$^{14}$ W/cm$^2$) and pulse length of 40 fs.}
\end{figure}

It is interesting to point out that the probability of excitation
in Fig.~\ref{fig:impro} shows distinct sharp peaks as a function
of R. To disentangle the source of these peaks, in
Fig.~\ref{fig:impro2} we examine the contributions to the $\pi_u$
excitation probability according to whether the return time t$_r$
falls within one, two, three or four optical cycles after the
tunneling electron is born. The excitation probabilities are
larger for returns within one or two optical cycles. Within the
first two optical cycles, the nuclear wave packets remain at small
R with small spreading and the returning electron has more kinetic
energy (see Fig.~\ref{fig:eng}). For the higher returns the
nuclear wave packet moves to larger R and spreads further  and the
smaller energies for the returning electron render the excitation
probabilities smaller.

We comment once again that with the inclusion of Coulomb
attraction on the motion of the rescattering electron, the maximum
returning electron energy is not given by 3.17U$_P$=29 eV for the
present peak intensity, but rather by 35 eV, as seen from
Fig.~\ref{fig:eng}. This has the effect of enhancing the
excitation to the $\pi_u$ state as well.

For peak laser intensity of 1.5 I$_0$ the results in
Fig.~\ref{fig:impro} show that direct impact ionization of D$_2^+$
by the rescattering electron is very small. The rescattering
mostly populates D$_2^+$ in the excited $\pi_u$ and $\sigma_u$
states. The dissociation of D$_2^+$ from an excited electronic
state would release a total kinetic energy given by
U(R$_0$)-U($\infty$), shared equally by D and D$^+$, respectively.
According to Fig.~\ref{fig:impro}, excitation by the rescattering
process peaks at characteristic internuclear separations related
to characteristic rescattering time t$_r$, thus measurement of the
D$^+$ fragment kinetic energies probes directly the recollision
times. This forms the basis of molecular clocks in the experiments
of Niikura {\it et~al.} \cite{Niikura02,Niikura03}. However, as
shown in Tong {\it et~al.} \cite{Tong03a} and in Alnaser {\it
et~al.} \cite{Alnaser03}, the excited D$_2^+$ ions are still in
the laser field and they can be further ionized by the lasers.
Thus we need to calculate the kinetic energy spectra of D$^+$
resulting from Coulomb explosion after these excited D$_2^+$ ions
are ionized by the laser.

\subsection{Field ionization of excited D$_2^+$ ion}
\begin{figure}[tb]
 \includegraphics[width=\figwidth]{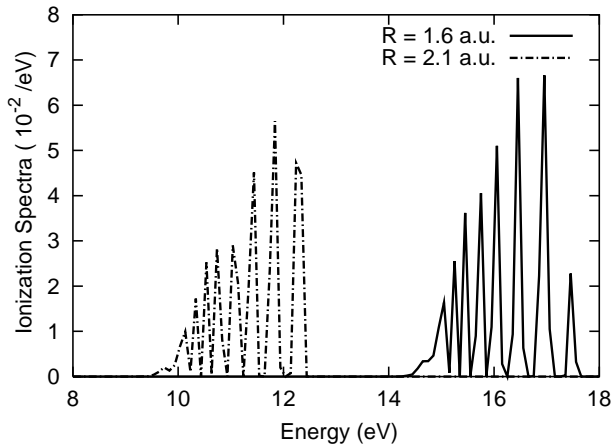}
 \caption{\label{fig:su_ion} Kinetic energy spectra of D$^+$ ion resulting
 from laser ionization of D$_2^+$ in the excited  $\sigma_u$
 state, for ion reaching the excited state initially at R=1.6 a.u.
 (solid curve) and R=2.1 a.u (dashed curve), respectively. Laser parameters: peak intensity
  1.5 x10$^{14}$ W/cm$^2$, pulse length 40 fs.
 }
\end{figure}

In this subsection we consider the ionization of D$_2^+$ from the
excited electronic states. We emphasize that we will consider peak
laser intensity within 0.5-5 I$_0$ only where rescattering is
important. In this intensity region,  D$_2^+$ is readily ionized
if it is in the $\pi_u$ excited state since its saturation
intensity is only about 0.1 I$_0$ because of  small ionization
energy. Thus we need only to calculate the ionization rate of
D$_2^+$ from the $\sigma_u$ state.  If the initial excitation to
$\sigma_u$ occurs at R, the total accumulated probability for
ionizing an electron by the laser field from the $\sigma_u$ state
is
\begin{eqnarray}{\label{eq:ionization}}
P_i (R,\infty) & = & 1-e^{-\int W(R') dt} \nonumber \\
       & = & 1-e^{-\int_{R}^{\infty} W(R')/v(R') dR'},\\
\hspace{-1cm} \mbox{with}\hspace{1cm} \frac{1}{2}\mu v^2(R') & = &
U(R)-U(R').
\end{eqnarray}
where $W(R')$ is the MO-ADK tunneling ionization rate described in
subsection A, $\mu$ is the reduced mass of the two nuclei, and
$U(R)$ is the total potential energy of the $\sigma_u$ state. The
$\sigma_u$ state created at $R$, followed by laser field
ionization at $R'$ will release a kinetic energy $E_i(R') =
U(R)-U(R')+1/R'$. Here we are more interested in the differential
ionization probability which is given by
\begin{eqnarray}
\frac{d P_i(R,R') }{dR'} = \frac{W(R')}{v(R')}e^{-\int_{R}^{R'}
W(R'')/v(R'') dR''},
\end{eqnarray}
or in terms of differential probability
per unit of kinetic energy
\begin{eqnarray}
\frac{d P_i(R,R') }{dE} & = & \frac{d P_i(R,R') }{d R'}
\frac{dR'}{d U}
\\ \frac{dR'}{d U} & = & \frac{1}{| \frac{ d U(R')}{dR'} | }
\end{eqnarray}
Fig.~\ref{fig:su_ion} shows the expected ionization spectra from
the $\sigma_u$
state if it is initially created at two different R's (R=1.6 and
2.1 a.u.), chosen to be the peak positions of the vibrational wave
packet at the first and the third returns. Clearly, the early
return releases more energy (higher energy peak). The spectra show
many sharp peaks since ionization occurs only when the laser field
is near its peak intensity at every half cycle.

To obtain the total ionization spectra, we need to add up
contributions from initial ionization at all values of R, i.e.,
\begin{eqnarray}\label{eq:ionP}
\frac{d P_{ion}}{d E} & = & \int \frac{d P_m}{d R} \frac{d
P_i(R,R') }{dE} dR
\end{eqnarray}
This integration is important primarily only for ionization from
the excited $\sigma_u$ state. For other excited electronic states,
due to the high ionization rate, ionization is complete within one
cycle or less and we can set $ R=R'$, and the differential
ionization spectra for these excited electronic  states are given
by
\begin{eqnarray}
\frac{d P_{ion}}{d E} & = &  \frac{d P_m}{d R}\frac{dR}{d U}.
\end{eqnarray}
The total ionization spectra are obtained by adding up
contributions from all the excited electronic states, and from the
initial ionization by the rescattering electron (very negligible).

For the dissociation process, the energy spectra are obtained from
\begin{eqnarray} \label{eq:disP}
\frac{d P_{dis}}{d E} & = & (1-P_i(R)) \frac{d P_m}{d
R}\frac{dR}{d U}.
\end{eqnarray}
The total dissociation spectra are obtained by adding up
contributions from all the excited electronic states. In reality,
the dissociation comes from the $\sigma_u$ excited state only. In
all other excited electronic states the D$_2^+$ ions are
immediately ionized by the laser in one optical cycle.

\section{Results and Discussion}\label{sec:results}

 The kinetic energy spectra of D$^+$ ions can be determined
without any coincidence, as in the experiments of Niikura {\it
et~al.} \cite{Niikura02,Niikura03}, or by detecting  the two D$^+$
ions in coincidence, as in the experiments of Staudte {\it et~al.}
\cite{Staudte02} or in Alnaser {\it et~al.} \cite{Alnaser03}. we
will present our simulation results for both types of experiments.

\subsection{Non-coincident D$^+$ kinetic energy spectra}
\begin{figure}[tb]
 \includegraphics[width=\figwidth]{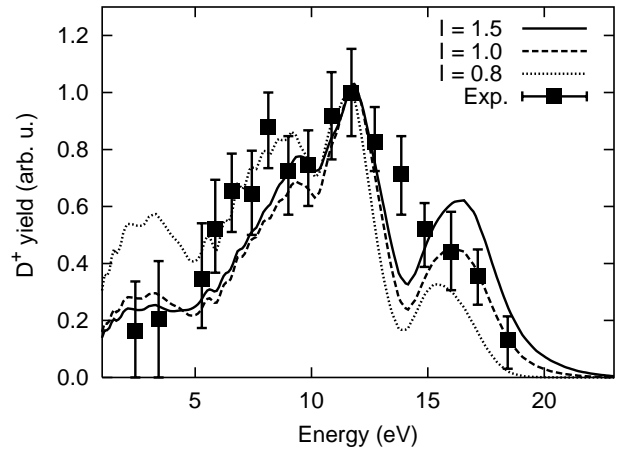}
 \caption{\label{fig:C_exp} D$^+$ yield at several laser
 intensities for a
 35 fs pulse length. The experimental data are from Ref.~\cite{Niikura03} for
 1.5 I$_0$ where I$_0$=10$^{14}$ W/cm$^2$. The peak values from
 experiment and from theory for 1.5 I$_0$ are normalized to each
 other. For peak intensities of 1.0 I$_0$ and 0.8 I$_0$, the
 yields
 have been multiplied by 1.4 and 3.0 respectively to have the same peak ion
 yield
 height.
  }
\end{figure}

In the experiments of Niikura {\it et~al.}
\cite{Niikura02,Niikura03}, the kinetic energy of D$^+$ ion was
measured in the direction perpendicular to the direction of laser
polarization. The measured D$^+$ signals come from ionization and
from dissociation. Thus,
\begin{eqnarray}\label{eq:non-co}
Signal \propto 2 \frac{d P_{ion}}{d E} + \frac{d P_{dis}}{d E}.
\end{eqnarray}
In Fig.~\ref{fig:C_exp} the experimental D$^+$ kinetic energy
spectra from Niikura {\it et~al.} \cite{Niikura03} are shown. The
energy
scale is the total breakup energy, or twice the energy of the
D$^+$ ion. The experiment was performed for a pulse of 40 fs and
peak intensity of 1.5 I$_0$. We have shown simulations with the
same laser parameters but with three peak laser intensities, at
1.5, 1.0 and 0.8 I$_0$. First we normalize the peak height at 12
eV between theory and experimental data at 1.5 I$_0$. Since the
peak positions do not vary with laser intensity, we can normalize
the calculated spectra at the two other intensities as well, with
a multiplicative factor of 1.4 and 3.0 for the 1.0 and 0.8 I$_0$,
respectively. If one compares the experimental spectra with the
theoretical one calculated at the same 1.5 I$_0$, clearly the high
energy peak near 16 eV from the theory is too high, while the
theoretical spectra between 5 and 10 eV are somewhat too low.
However it appears that the discrepancy can be reconciled if one
takes into account of the volume effect in that the experimental
spectra have to be integrated over a volume where the intensities
are less than the peak value. The energy resolution and the finite
acceptance angles can all contribute to the smoother experimental
spectra. One of course should also take this ``better agreement''
with caution in view that the peak intensity of the laser is often
not known precisely.

\begin{figure}[b]
 \includegraphics[width=\figwidth]{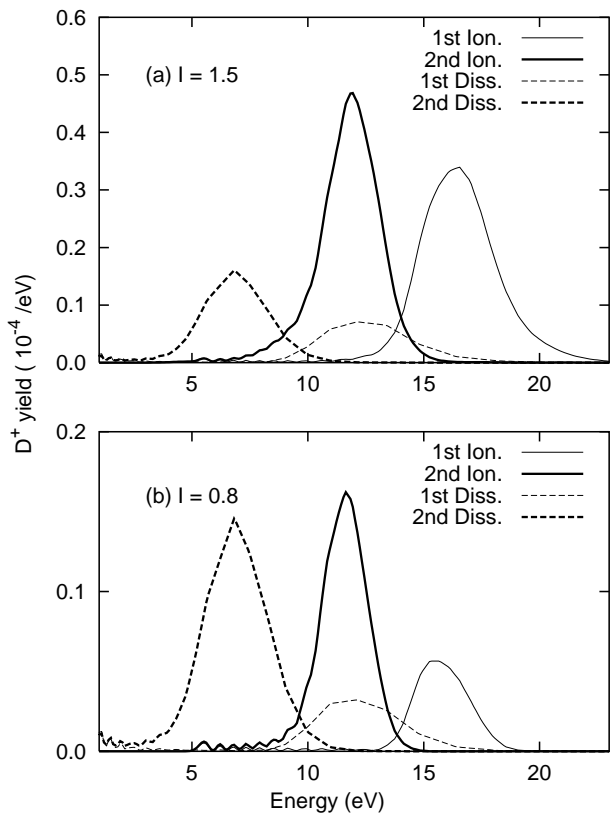}
 \caption{\label{fig:res1a} Decomposition of D$^+$ ion yields into
 contributions from dissociation and ionization, and for rescattering
 occurring within the first and the second optical cycle after
 the initial tunneling ionization.  The peak laser intensities are
 (a) I = 1.5 I$_0$ and (b) I = 0.8 I$_0$, where
 I$_0$=10$^{14}$ W/cm$^2$ and pulse length is 40 fs.}
\end{figure}

  One of the major goals of the simulation is to unravel the origin
of the structure in the kinetic energy spectra which in turn would
provide insight of the working of the molecular clock. For this
purpose, we show in Fig.~\ref{fig:res1a} the calculated kinetic
energy spectra, but separate the contributions from dissociation
and ionization, and from rescattering occurred after one or two
optical cycles, or equivalently, from the first (t$_1$) or the
third returns(t$_3$), at two laser intensities, 1.5 and 0.8 I$_0$.
At the higher intensity, in this figure we notice: (1) ionization
is much stronger than dissociation; (2) the peak from the third
return (2nd cycle) is higher than from the first return; (3) The
width of the peak from the first return is broader than the peak
from the third return. The broadening is a consequence of the
factor $dR/dU$ in Eqs.~(\ref{eq:ionP}) and (\ref{eq:disP}) which
is approximately given by R$^2$. Another interesting observation
is that the peak position of the dissociation spectra from the
first return almost coincides with the peak position in the
ionization spectra from the third return. This shift is due to the
binding energy of the excited electronic states.

In Niikura {\it et~al.}'s experiment \cite{Niikura03} the peak at
12 eV was attributed to originate from the dissociation of D$_2^+$
via the $\sigma_u$ curve at the first return. In other words, this
peak reads the clock at t$_1$. According to our simulation, the
peak comes from ionization following rescattering  at the third
return, and this peak should read the clock at t$_3$.

Contributions to the D$^+$ signal from dissociation do become more
important at   lower laser intensity, as shown in
Fig.~\ref{fig:res1a}(b). Even at this intensity, the peak at 12 eV
still comes mostly from the ionization following rescattering at
t$_3$ instead of  dissociation following rescattering at t$_1$.
Furthermore the third return peak is higher than the first return
peak for either dissociation or ionization. We remark that the
spectra in Fig.~\ref{fig:C_exp} were calculated including
contributions up to four or five optical cycles after the initial
tunneling ionization and convergence of the calculation was
checked.

\subsection{D$^+$ coincident kinetic energy spectra}

\begin{figure}[tb]
 \includegraphics[width=\figwidth]{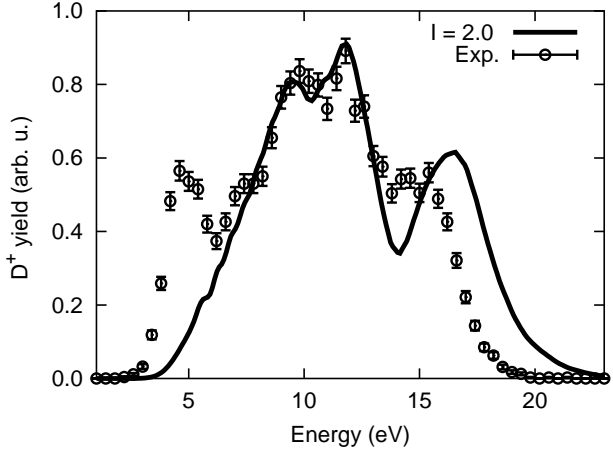}
 \caption{\label{fig:L_exp} Comparison of D$^+$ ion spectra resulting from the
 double ionization of D$_2$ molecules in a laser field.
   The experiment data
 are from Ref.~\cite{Alnaser03} for peak laser intensity of 2.8 I$_0$ and the
 theoretical simulation is for laser peak intensity of 2I$_0$, where
 I$_0$=10$^{14}$ W/cm$^2$ and the pulse length is 35 fs.}
\end{figure}

The D$^+$ ion kinetic energy distributions in laser-D$_2$
interactions have been determined in coincidence measurements
where the two D$^+$ ions were detected simultaneously by Staudte
{\it et~al.} \cite{Staudte02} and more recently by Alnaser {\it
et~al.} \cite{Alnaser03}. In the latter experiment, the branching
ratios of ionization with respect to dissociation had been
measured as well, for peak laser intensities of 1-5 I$_0$. Their
data for peak intensity of 2.8 I$_0$ are shown in
Figure~\ref{fig:L_exp}.
The experiment used a 35 fs pulse with mean wavelength of 800 nm.
The D$^+$ spectra are from Coulomb explosion of ions at
60-80$^\circ$ with respect to the direction of the linear
polarization of the laser field. In the figure we show the result
of our theoretical simulation for laser intensity of 2.0 I$_0$. We
found best overall agreement with the experimental data at this
intensity without considering volume effect and the fact that the
theoretical calculation was carried out for molecules aligned
perpendicular to the laser polarization while the experiments
measured ions coming out of 60-80$^\circ$ with respect to the
laser polarization. The simulated spectra near the kinetic energy
peak region of 7-12 eV agree quite well with the data, but the
peak near 17 eV is more pronounced in the simulation.

\begin{figure}[tb]
 \includegraphics[width=\figwidth]{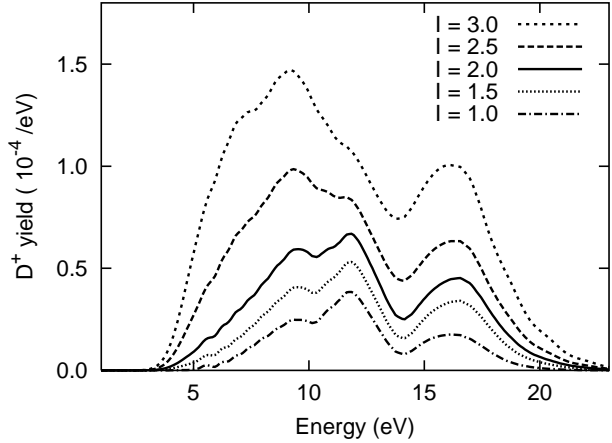}
 \caption{\label{fig:L_exp1} Simulated D$^+$ ion yield from the double ionization
 of D$_2$ by the rescattering process at several peak
 laser intensities in units of I$_0$=10$^{14}$ W/cm$^2$. The pulse length is 35 fs.}
\end{figure}

\begin{figure}[b]
 \includegraphics[width=\figwidth]{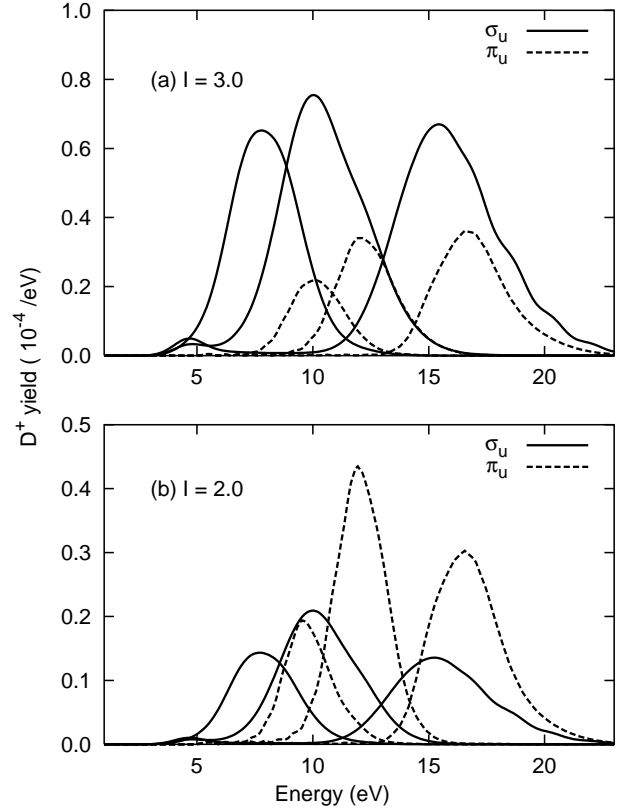}
 \caption{\label{fig:L_exp2} D$^+$ yield from laser ionization
 via the $\sigma_u$ (solid line) and $\pi_u$ (dashed line)
 excited states with laser intensity (a) I = 3.0 I$_0$ and (b)
 I = 2.0 I$_0$ (I$_0$=10$^{14}$ W/cm$^2$). Each yield is further
 decomposed into contributions for rescattering occurring after
 one, two and three optical cycles. The laser pulse length is 35 fs.}
\end{figure}

A direct comparison of simulated kinetic energy spectra with
experimental data is complicated in general not only by the volume
effect, the angular resolution of the D$^+$ product, but also the
difficulty of knowing the peak laser intensity  precisely. In
Fig.~\ref{fig:L_exp1} we show the  yield for making two D$^+$ ions
vs the total
kinetic energy for peak laser intensity from 1.0-3.0 I$_0$. The
calculations were done for 35 fs pulse and mean wavelength of
800~nm and with molecules aligned perpendicular to the laser
polarization direction. Clearly the yield increases rapidly with
laser intensity. We further note that the peak  positions in the
spectra do change with laser intensity. In particular, the main
peak shifts to  lower kinetic energy at higher laser  intensity.
To understand the reason of this shift, in Fig.~\ref{fig:L_exp2}
we separate the kinetic energy peaks into contributions from the
$\sigma_u$ and from the $\pi_u$ curves, and for rescattering
occurring after one, two and three optical cycles following
tunneling ionization. Recall that we consider D$^+$ from
ionization only here. At 2.0 I$_0$, we note the larger
contribution comes mostly from ionization of D$_2^+$ in the
$\pi_u$ state, although contribution from $\sigma_u$ is not
negligible. From Fig.~\ref{fig:L_exp2}(b) one can clearly identify
the two peaks approximately at 10 and 12 eV in
Fig.~\ref{fig:L_exp} can be attributed to ionization from
$\sigma_u$ and $\pi_u$, respectively, for rescattering collision
from the third return. At 3.0 I$_0$ (Fig.~\ref{fig:L_exp2}(a) ),
due to the larger contribution from the $\sigma_u$ excited state,
the peak positions in the kinetic energy spectra are shifted to
lower values. Thus the sum kinetic energy spectra at the two
higher intensities look different from those at lower intensities,
as seen in Fig.~\ref{fig:L_exp1}. Fig.~\ref{fig:L_exp2}(a) also
shows that contribution from the third cycle  becomes relatively
more important at higher intensity. At higher intensity, the
rescattered electron has larger kinetic energy. Thus it takes more
time for the Coulomb attraction to bring the electron to come near
the ion for the rescattering to occur.

Figure~\ref{fig:L_exp2} also  illustrates how the working
condition for using rescattering model to measure the precise time
in a molecular clock can be limited. The kinetic energy spectra
from each excited electronic state of D$_2^+$ has  relatively well
specified and distinct peak positions from the first, 3rd and 5th
returns. Such peak positions immediately give information about
the molecular clock since each peak position does not depend on
the laser intensity. However, when ionization from $\sigma_u$
channel also contributes then the combined sum would shift the
peak positions as the laser intensity is changed, as shown in
Fig.~\ref{fig:L_exp1}. Thus to read the molecular clock
accurately, one has to choose laser intensity where only one of
the excited D$_2^+$ electronic state contributes mostly to the
ionization signal. Failure to do so would compromise the accuracy
of the clock. Since the relative contributions of the ionization
signals from $\sigma_u$ and $\pi_u$ are expected to change with
laser intensity and with the alignment of the molecules, this also
helps explain why the valleys in the experimental spectra are
usually less sharply peaked than the ones simulated from the
theory at a given peak laser intensity.

\subsection{Laser- H$_2$ interactions and wavelength dependence}
Clearly the present method can be used to predict the kinetic
energy spectra if H$_2$ is used as the target. The only difference
in H$_2$ is that it has smaller reduced mass such that the wave
packet propagates faster, and thus kinetic energy spectra will be
shifted to lower energies. If the wavelength of the laser is
increased, the period is longer and thus the kinetic energy
spectra also will shift to lower energies.  We have applied the
present theoretical model to study the comparison of kinetic
energy spectra taken for H$_2$ and D$_2$ simultaneously
\cite{Alnaser03}, and also the variation of the kinetic energy
spectra when  wavelength was varied as in the experiment of
Niikura {\it et~al.} \cite{Niikura03}, see Tong {\it et~al.}
\cite{Tong03a}.

\section{Summary and Conclusion}\label{sec:conclusion}

In this paper we have provided a comprehensive study of the
elementary processes of the rescattering mechanism leading to the
fragmentation of D$_2^+$ following the initial tunneling
ionization of a D$_2$ molecule in a short intense laser pulse.
Ionization rates of D$_2^+$ from the excited electronic states and
impact excitation and ionization cross sections by the returning
electron have been obtained based on the MO-ADK theory and from
semi-empirical formulation, respectively. Following the initial
idea of Corkum and coworkers, we showed that the kinetic energy
spectra of D$^+$ in the higher energy region (5 to 10 eV per D$^+$
ion) can be used as a molecular clock which can be read with
subfemtosecond accuracy. Through our detailed simulation, we
concluded that the dominant peak in the D$^+$ kinetic energy
spectrum is due to the further ionization of the excited D$_2^+$
following impact excitation by the returning electron, and this
excitation occurs not at the first return but mostly at the third
return. We have compared our simulation results with the recent
experiments of Niikura {\it et~al.\ } and of Alnaser {\it et~al.\
} with general good agreement. Further experimental studies in
terms of dependence on laser wavelength, pulse duration and
alignment angles may provide more critical test on the present
theoretical model. From the theoretical viewpoint, despite of the
semi-empirical nature of the present modeling, we do not expect
any meaningful pure {\it ab initio\ } quantum calculations viable
in the foreseeable future. The present model has the further
advantage that the mechanism for producing each individual peaks
in the kinetic energy spectra can be identified and the effect of
laser parameters can be readily tested. On the other hand, the
semi-empirical nature of the modeling can claim its reliability
only after it has been exposed to more stringent tests from the
experiment.

\acknowledgments

This work is in part supported by Chemical Sciences, Geosciences
and Biosciences Division, Office of Basic Energy Sciences, Office
of Science, U. S. Department of Energy. CDL also wishes to thank
Igor Bray for communicating to him the partial 1s
$\rightarrow$2p$_m$ (m=0,1) cross sections.


\begin{thebibliography}{10}

\bibitem{Sandig00}
K. Sandig, H. Figger, and T.~W. Hansch, Phys. Rev. Lett. {\bf 85},
4876
  (2000).

\bibitem{Staudte02}
A. Staudte {\it et~al.}, Phys. Rev. A {\bf 65},  020703(R)
(2002).

\bibitem{Niikura02}
H. Niikura {\it et~al.}, Nature {\bf 417},  917  (2002).

\bibitem{Niikura03}
H. Niikura {\it et~al.}, Nature {\bf 421},  826  (2003).

\bibitem{Sakai03}
H. Sakai {\it et~al.}, Phys. Rev. A {\bf 67},  063404  (2003).

\bibitem{Posthumus99}
J.~H. Posthumus {\it et~al.}, J. Phys. B {\bf 32},  L93  (1999).

\bibitem{Codling89}
K. Codling, L.~J. Frasinski, and P.~A. Hatherly, J. Phys. B {\bf
22},  L321
  (1989).

\bibitem{Seideman95}
T. Seideman, M.~Y. Ivanov, and P.~B. Corkum, Phys. Rev. Lett. {\bf
75},  2819
  (1995).

\bibitem{Zuo95a}
T. Zuo and A.~D. Bandrauk,  {\bf 52},  R2511  (1995).

\bibitem{Constant96}
E. Constant, H. Stapelfeldt, and P.~B. Corkum, Phys. Rev. Lett.
{\bf 76},  4140
   (1996).

\bibitem{Suzor95}
A. Giusti-Suzor {\it et~al.}, J. Phys. B {\bf 22},  309  (1995).

\bibitem{Codling93}
K. Codling and L.~J. Frasinski, J. Phys. B {\bf 26},  783  (1993).

\bibitem{Bandrauk99b}
A. Bandrauk, Comments At. Mol. Phys. {\bf 1(3) D},  97  (1999).

\bibitem{Fittinghoff92}
D. Fittinghoff, P. Bolton, B. Chang, and K. Kulander, Phys. Rev.
Lett. {\bf
  69},  2642  (1992).

\bibitem{Kondo93a}
K. Kondo {\it et~al.}, Phys. Rev. A {\bf 48},  R2531  (1993).

\bibitem{Corkum93}
P.~B. Corkum, Phys. Rev. Lett. {\bf 71},  1994  (1993).

\bibitem{Walker94a}
B. Walker {\it et~al.}, Phys. Rev. Lett. {\bf 73},  1227  (1994).

\bibitem{Brabec96}
T. Brabec, M.~Y. Ivanov, and P.~B. Corkum, Phys. Rev. A {\bf 54},
R2551
  (1996).

\bibitem{Sheey98}
B. Sheehy {\it et~al.}, Phys. Rev. A {\bf 58},  3942  (1998).

\bibitem{Becker00}
A. Becker and F.~H.~M. Faisal, Phys. Rev. Lett. {\bf 84},  3546
(2000).

\bibitem{Kopold00a}
R. Kopold, D.~B. Milosevic, and W. Becker, Phys. Rev. Lett {\bf
84},  3831
  (2000).

\bibitem{Yudin01}
G.~L. Yudin and M.~Y. Ivanov, Phys. Rev. A {\bf 63},  033404
(2001).

\bibitem{Yudin01a}
G.~L. Yudin and M.~Y. Ivanov, Phys. Rev. A {\bf 64},  035401
(2001).

\bibitem{Fu01}
L.-B. Fu, J. Liu, J. Chen, and S.-G. Chen, Phys. Rev. A {\bf 63},
043416
  (2001).

\bibitem{Tong02b}
X.~M. Tong, Z.~X. Zhao, and C.~D. Lin, Phys. Rev. A {\bf 66},
033402  (2002).

\bibitem{Alnaser03}
A. Alnaser {\it et~al.}, Phys. Rev. Lett.  (2003), (in press).

\bibitem{Perelomov66}
A.~M. Perelomov, V.~S. Popov, and M.~V. Terentev, Zh. Eksp. Teor.
Fiz. {\bf
  50},  1393  (1966), [Sov. Phys. JETP {\bf 23}, 924 (1966)].

\bibitem{Ammosov86}
M.~V. Ammosov, N.~B. Delone, and V.~P. Krainov, Zh. Eksp. Teor.
Fiz. {\bf 91},
  2008  (1986), [Sov. Phys. JETP {\bf 64},1191 (1986)].

\bibitem{Saenz00a}
A. Saenz, Phys. Rev. A {\bf 61},  051402(R)  (2000).

\bibitem{Chu01a}
X. Chu and S.~I. Chu, Phys. Rev. A {\bf 63},  013414  (2001).

\bibitem{Bray}
I. Bray, CCC-database  ,
\url{http://atom.murdoch.edu.au/CCC-WWW/index.html}.

\bibitem{Kim00}
Y.~K. Kim, K.~K. Irikura, and M.~A. Ali, J. Res. NIST {\bf 105},
285  (2000).

\bibitem{Bray03}
I. Bray,   (2003), private communication.

\bibitem{Merabet99}
H. Merabet {\it et~al.}, Phys. Rev. A {\bf 60},  1187  (1999).

\bibitem{Peek64}
J.~M. Peek, Phys. Rev. {\bf 134},  A877  (1964).

\bibitem{Tong03a}
X.~M. Tong, Z.~X. Zhao, and C.~D. Lin, Phys. Rev. Lett.  (2003),
(in press).

\end{thebibliography}

\end{document}